\documentclass[sigconf]{acmart} 

\usepackage{balance}
\usepackage{booktabs} 
\usepackage[vlined,figure]{algorithm2e}
\usepackage{amssymb}
\usepackage{amsmath}

\usepackage{listings}
\newcommand\lt[1]{{\lstinline+#1+}}
\definecolor{darkblue}{rgb}{0,0,0.5}
\definecolor{dkgreen}{rgb}{0,0.5,0}
\definecolor{dkred}{rgb}{0.5,0,0}
\definecolor{gray}{rgb}{0.5,0.5,0.5}
\definecolor{dkgray}{rgb}{0.3,0.3,0.3}
\usepackage{tikz}
\usetikzlibrary{arrows,shadows,positioning,shapes,automata}
\tikzset{
  arrow_to/.style = {->,shorten <= 1pt, shorten >= 1pt, >=stealth',semithick},
  mycircle/.style = {circle,thick,fill=blue!10,draw=blue!80,
    align=center
  },
  myellipse/.style = {ellipse,thick,fill=blue!10,draw=blue!80,
    align=center
  },
  myrect/.style = {rectangle,thick,fill=red!10,draw=red!80},
  myroundrect/.style = {myrect,
    rounded corners=0.5em,
    align=center 
  },
  myline/.style = {thick, draw=red},
  mygeoshape/.style = {myline,fill=red!5}
}

\newcommand{\tool}{\textsf{NumInv}}

\newif\ifcopyright
\copyrightfalse

\ifcopyright
\setcopyright{acmcopyright}
\else
\setcopyright{none}
\fi

\acmDOI{10.1145/3106237.3106281}
\acmISBN{978-1-4503-5105-8/17/09}
\acmConference[ESEC/FSE'17]{2017 11th Joint Meeting of the European Software Engineering Conference and the ACM SIGSOFT Symposium on the Foundations of Software Engineering}{September 4--8, 2017}{Paderborn, Germany} 
\acmYear{2017}
\copyrightyear{2017}
\acmPrice{15.00}

\begin{document}

\author{ThanhVu Nguyen}
\affiliation{%
  \institution{University of Nebraska-Lincoln, USA}
}

\author{Timos Antonopoulos}
\affiliation{%
  \institution{Yale University, USA}
}
\author{Andrew Ruef}
\affiliation{%
  \institution{University of Maryland, USA}
}
\author{Michael Hicks}
\affiliation{%
  \institution{University of Maryland, USA}
}

\title{Counterexample-Guided Approach to Finding \\
Numerical Invariants
}

\begin{abstract}
Numerical invariants, e.g., relationships among numerical variables in a program, represent a useful class of properties to analyze programs.
General polynomial invariants represent more complex numerical relations, but they are often required in many scientific and engineering applications.
We present {\tool}, a tool that implements a counterexample-guided invariant generation (CEGIR) technique to automatically discover numerical invariants, which are  polynomial equality and inequality relations among numerical variables.
This CEGIR technique infers candidate invariants from program traces and then checks them against the program source code using the KLEE test-input generation tool.
If the invariants are incorrect KLEE returns counterexample traces, which help the dynamic inference obtain better results.
Existing CEGIR approaches often require sound invariants, however {\tool} sacrifices soundness and produces results that KLEE cannot refute within certain time bounds.
This design and the use of KLEE as a verifier allow {\tool} to discover useful and important numerical invariants for many challenging programs.

Preliminary results show that {\tool} generates required invariants for understanding and verifying correctness of programs involving complex  arithmetic.
We also show that {\tool} discovers polynomial invariants that capture precise complexity bounds of programs used to benchmark existing static complexity analysis techniques.
Finally, we show that {\tool} performs competitively comparing to state of the art numerical invariant analysis tools.
\end{abstract}

\ccsdesc[500]{Software and its engineering~Software verification and validation}
\ccsdesc[500]{Software and its engineering~Software verification}
\ccsdesc[500]{Software and its engineering~Automated static analysis}
\ccsdesc[500]{Software and its engineering~Dynamic analysis}

\keywords{Dynamic and Static Invariant Analyses, Counterexample-guided Algorithms, Numerical Domains, Program and Correctness Analyses, Test-input Generation}

\maketitle

\section{Introduction}\label{sec:intro}
\newcommand{\Astree}{Astr\'ee}

The automated discovery of \emph{program invariants}---relations among
variables that are guaranteed to hold at certain locations of a
program---is an important research area in program analysis and
verification. Generated invariants can be used to prove correctness
assertions, reason about resource usage, establish security
properties, provide formal documentation, and more~\cite{slam,blast,ESP,z3ms,leroy06,daikon}.

A particularly useful class of invariants are \emph{numerical
  invariants}, which involve relations among numerical program
variables.  Within this class of invariants, \emph{nonlinear polynomial}
relations, e.g., $x \le y^2 , x = qy + r$, arise in many scientific,
engineering, and safety- and security-critical applications.\footnote{We refer to nonlinear polynomial relations such as $x=qy+r, x \le y^2$ simply as \emph{polynomial relations}.}  For
example, the commercial analyzer {\Astree}, which has been applied to
verify the absence of errors in the Airbus A340/A380 avionic
systems~\cite{CCF05,BCC03}, implements the ellipsoid abstract domain~\cite{feret2004static}
to represent and analyze a class of quadratic inequality
invariants. 
Complexity analysis, which aims to determine a program's
performance
characteristics~\cite{pldi09,Gulwani:2009:SSC:1575060.1575069,Hoffmann:2012:RAM:2362216.2362294},
perhaps to identify possible security problems~\cite{hoffman17verifying, AntonopoulosGHK17}, also makes use of polynomial
invariants, e.g., $O(n^2 + 2m)$ where $n,m$ are some program inputs.
In addition, such polynomial invariants have been found useful in the
analysis of hybrid systems~\cite{RFM05,SSM05}, and in fact are
required for implementations of common mathematical functions such as
\textsf{mult, div, square, sqrt} and \textsf{mod}.

Numerical  invariants can be discovered via static
and dynamic program analyses. A static analysis can reason about all
program paths soundly, but doing so is often expensive and is only
possible for relatively simple forms of invariants~\cite{vuphdthesis}.
Dynamic analyses limit their attention to only some of a program's
paths, and as a result can often be more efficient and produce more
expressive invariants, but provide no guarantee that those invariants are
correct~\cite{daikon,tosem2013}. Recently, several systems (such as
PIE~\cite{Padhi:2016:DPI:2908080.2908099},
ICE~\cite{Garg:2016:LIU:2837614.2837664} and {\it Guess-and-Check}~\cite{sharma2013data}) have been developed that
take a hybrid approach: use a dynamic analysis to infer
\emph{candidate invariants} but then confirm these invariants are
correct for all inputs using a \emph{static verifier}. When invariants
are incorrect the verifier returns counterexample traces which the
dynamic inference engine can use to infer more accurate
invariants.  This iterative process is called \emph{CounterExample
  Guided Invariant geneRation} (CEGIR).

\begin{figure}
  \begin{minipage}{0.48\linewidth}
\begin{lstlisting}[numbers=none,mathescape,xleftmargin=0.2cm,emph={L1,L2}]
int cohendiv(int x, int y){
  assert(x>0 && y>0);
  int q=0; int r=x;
  while(r $\ge$ y){
    int a=1; int b=y;
    while[L1](r $\ge$ 2*b){
      a = 2*a; b = 2*b;
    }
    r=r-b; q=q+a;
  }
  [L2]
  return q;
}
\end{lstlisting}
\end{minipage}
\hfill
\begin{minipage}{0.48\linewidth}
\centering
\begin{tabular}{c  c c  c  c c}
\multicolumn{6}{r}{\underline{Traces}:}\\
\\
$x$&$y$&$a$&$b$&$q$&$r$\\
\midrule
15& 2& 1& 2& 0& 15\\
15& 2& 2& 4& 0& 15\\
15& 2& 1& 2& 4& 7\\
\multicolumn{3}{c}{}&\multicolumn{1}{c}{$\vdots$}&\multicolumn{1}{c}{}\\
\midrule            
4& 1& 1& 1& 0& 4\\
4& 1& 2& 2& 0& 4\\
\multicolumn{3}{c}{}&\multicolumn{1}{c}{$\vdots$}&\multicolumn{1}{c}{}\\
\end{tabular}
\end{minipage}
\caption{An integer division program and example trace values at
  location $L1$ on inputs $(x=15, y=2)$ and $(x=4,y=1)$. Among other
  invariants, two key loop invariants discovered at \textsf{L1} are
  $b = ya$ and $x = qy+r$, with the latter also found as the postcondition
  at \textsf{L2}.}
\label{fig:motiv}
\end{figure}

While the CEGIR approach is promising, existing tools have some
practical limitations. One limitation is that they find 
invariants strong enough to prove a particular (programmer-provided)
postcondition where the quality of the generated invariants depends on the
strength of the postcondition.  As such, they are not well suited for
automated analyses on code that lacks such formal specifications. Another
limitation is that these tools employ a \emph{sound} static verifier,
which aims to definitively prove that an invariant holds. While this
is a good goal, it turns out to be a significant restriction on the
quality of the invariants that can ultimately be inferred---it
can be quite challenging to do when invariants are nonlinear polynomials and
involve many program variables. For example, consider the program in
Figure~\ref{fig:motiv}, which implements Cohen's algorithm for integer
division~\cite{cohen1}. Two important loop invariants (at \textsf{L1})
are $b = ya$ and $x = qy+r$, as they both point directly to the
correctness of the algorithm.\footnote{$x = qy+r$ describes the intended
behavior of integer division: the dividend $x$ equals the divisor $y$
times the quotient $q$ plus the remainder $r$.} Neither PIE nor ICE
can infer these invariants (both tools time out).
  
In this paper we present a new CEGIR algorithm called
{\tool} that overcomes these
limitations. It has two main components. First, it uses algorithms
from DIG~\cite{vuicse2012,tosem2013} to dynamically infer expressive
polynomial equality invariants and linear inequality relations from
traces at specified program locations. Second, it uses
KLEE~\cite{cadar2008klee}, a symbolic executor, to check candidate
invariants and produce counterexamples when they fail to hold. To
check that an invariant $p$ holds at location \textsf{L}, {\tool} transforms the
input program so that \textsf{L} is guarded by the conditional $\neg{p}$. If
KLEE is able to reach \textsf{L} then $p$ must not be an invariant, and so it
outputs a counterexample consisting of the relevant input values at
that location. On the other hand, if KLEE never reaches that location prior
to timing out, then {\tool} accepts the invariant as correct. 
Although this technique is unsound, KLEE, by its nature as a symbolic
executor, turns out to be very effective in discovering counterexamples 
to refute invalid candidates.

For the example in Figure~\ref{fig:motiv}, {\tool} is able to find the
critical equalities mentioned above, along with many other
useful inequalities. 
These invariants help understand the precise semantics of the program and verify its correctness properties.
Moreover, by instrumenting the program with a counter
variable, {\tool} can even infer program
running times as a function of the inputs. For example,
{\tool} is able to infer the precise running time of the program in
Figure~\ref{fig:triple} (page~\pageref{fig:triple}) which has 
a tricky, triple-nested loop.

We evaluated {\tool} by using it to infer invariants on more than 90
benchmark programs taken from the NLA~\cite{vuicse2012} and
HOLA~\cite{Dillig:2013:IIG:2509136.2509511} suites for program 
verification and from examples in the literature on complexity bound
analysis~\cite{pldi09,Gulwani:2009:SSC:1575060.1575069,DBLP:conf/popl/GulwaniMC09}. 
Our results show that 
{\tool} generates sufficiently strong invariants to verify
correctness and to understand the semantics of 23/27 NLA programs
containing nontrivial arithmetic and polynomial relations.  We also
find that {\tool} discovers highly precise invariants describing nontrivial complexity bounds for 18/19 programs used to
benchmark static complexity analysis techniques (in fact, for 4
programs, {\tool} obtains more informative bounds than what were given
in the literature).  We note that both ICE and PIE cannot find any of
these invariants produced by {\tool}, even when we
explicitly tell these tools that they should attempt to verify these
invariants.  Finally, on the 46 HOLA programs, we compare {\tool}
directly with PIE. We find it performs competitively: in 36/46 cases
its inferred invariants match PIE's, are stronger, or are more descriptive.

Thus, although {\tool} can potentially return unsound invariants,
our experience shows that it is practical and effective in removing
invalid candidates and in handling difficult programs with complex
invariants.  We believe that {\tool} strikes a practical balance
between correctness and expressive power, allowing it to discover
complex, yet interesting and useful invariants out of the reach of the
current state of the art.

\section{Overview}\label{sec:overview}

{\tool} generates invariants using the technique of
\emph{counterexample-guided invariant generation} (CEGIR). 
At a high level, CEGIR consists of two components: a \emph{dynamic analysis}
that infers candidate invariants from execution traces, and a
\emph{static verifier} to check candidates against the program code.
If a candidate invariant is spurious, the verifier also provides
counterexamples (\emph{cex}s). Traces from these cexs are recycled to
repeat the process, hopefully producing accurate results. These steps
of inferring and checking repeat until no new cexs or (true) invariants
are found. The CEGIR approach is basically exploiting the observation
that inferring a sound solution directly is often harder than checking
a (cheaply generated) candidate solution.  

Other promising CEGIR algorithms, e.g., the ICE, PIE and {\it Guess-and-Check} tools, have been developed in recent years that take the same
approach~\cite{Garg:2016:LIU:2837614.2837664, Padhi:2016:DPI:2908080.2908099, sharma2013data}, though they refer to it differently. In
particular they refer to CEGIR as a \emph{data
  driven} or  \emph{black-box} approach, where the dynamic analysis is called the
\emph{student} or \emph{learner}, and the static verifier is called
the \emph{teacher} or \emph{oracle.} These approaches have been able
to prove correctness of specifications by inferring inductive loop invariants, or sufficient and necessary preconditions.
Some of these works (ICE and PIE) are verification oriented, i.e. they infer invariants to specifically prove a given assertion.
In this approach, the computation of these ``helper'' invariants strictly depends on the given assertions, e.g., if the intended assertion is \textsf{True} then the inferred invariant can be just \textsf{True}.
We review these works in more detail in Section~\ref{sec:related}.

{\tool} has different goals and takes a different approach.
Our goals are both discovery and verification, and our approach is to find the strongest possible invariant at any arbitrarily given location.
When given an undocumented program, {\tool} can discover interesting properties and provide formal specifications.
For example, {\tool} can reveal a stronger postcondition than the user might think to write down, and the user doesn't have to write down any postconditions at all. 
Moreover, when given a specific assertion, the resulting invariant from {\tool} can help prove it (e.g., if the invariant matches or is stronger than the assertion).
Empirically, {\tool} can  frequently infer invariants that are at least as strong as the postcondition, and frequently, stronger. 

\subsection{{\tool}}

{\tool} infers candidate invariants using the algorithms from
DIG~\cite{vuicse2012,tosem2013}, which produce equality and inequality
relations from traces. To check invariants, {\tool} invokes
KLEE~\cite{cadar2008klee}, a symbolic executor that is able to
synthesize test cases for failing tests. 

\paragraph{KLEE as a ``verifier''}

{\tool} generates candidate invariants at
program location \textsf{L} of interest (e.g., at the start of loops or at
the end of functions).
To check whether a property $p$ holds at
a location \textsf{L}, {\tool} asks KLEE to determine the
reachability of the location \textsf{L} when guarded by $\neg p$. For
example, to check whether the relation $x = qy+r$ is an
invariant at some location \textsf{L}, {\tool} modifies the program
as follows
\begin{lstlisting}[numbers=none,mathescape,xleftmargin=0.5cm,emph={L},morekeywords={save}]
  ...
  if (!(x==qy+r)){
    [L]
    save(x,y,q,r); //cex traces
    abort(); 
  }
  ...
\end{lstlisting}
KLEE then runs this program, systematically exploring the space of
possible inputs. If, during this process, location \textsf{L} is
reached, then the relation does not hold, so a cex consisting of the
values of the relevant input variables is saved for subsequent inference.
On the other hand, KLEE may be able to explore all program paths and thus verify that indeed that invariant $p$ holds.
Or, if this is infeasible, {\tool} terminates KLEE after some timeout.

The use of KLEE as the verifier is a key feature of {\tool}. Because
programs often contain a very large number of possible paths, KLEE
rarely explores all of them. However, in our experience
(Section~\ref{sec:results}), if it does not quickly find a counterexample
for $p$ then $p$ very likely holds. This is true even when $p$
is a nonlinear polynomial relation. As such, KLEE serves as a
practical improvement over existing theorem provers and constraint
solvers, for which reasoning over general polynomial arithmetic is a
significant challenge.

\paragraph{Inferring polynomial equalities and linear inequalities}

{\tool} uses two CEGIR algorithms to find candidate numerical
relations $p$ at program locations of interest. The first algorithm
finds \emph{polynomial equalities}.  To do this, for each program location
\textsf{L}, {\tool} produces a \emph{template} equation
$c_1t_1 + c_2t_2 \dotsb c_nt_n = 0$. This equation contains $n$
unknown coefficients $c_i$ and $n$ \emph{terms} $t_i$, with one term
for each possible combination of relevant program variables, up to
some degree $d$. {\tool} calls KLEE on the 
program to systematically obtain many possible valuations of relevant
variables at \textsf{L}. Each distinct observed valuation, which we
call a \emph{trace}, is substituted into the template to form an
instantiated equation. After obtaining at least $n$ traces, {\tool}
solves the $c_i$ using the resulting set of equations. Substituting
the solutions back into the template, we can extract candidate
invariants. At this point, {\tool} enters a CEGIR
loop that tests the candidate invariants by using KLEE as described
above. Any spurious invariants are dropped, and the corresponding cex
traces are used to infer new candidates, as described above, until no
additional true invariants are found. 

\tool's second algorithm tries to infer \emph{linear inequalities} in
the form of \emph{octagons}, which are inequalities over two
variables, containing eight edges. It refines the bounds on the
candidate invariants using a divide-and-conquer algorithm. Once again,
{\tool} estimates and obtains an initial set of traces. It enumerates
all possible octagonal inequality forms involving one and two
variables and uses KLEE to check inequalities under these forms are
within certain ranges $[minV, maxV]$. It then narrows this range,
iteratively seeking tighter lower and upper bounds.

Finally, from the obtained equality and inequality invariants, {\tool} removes any
invariants that are logical implications of other invariants.  For
instance, we suppress the invariant $x^2 = y^2$ if another invariant
$x = y$ is also found because the latter implies the former. We check
possible implications using an SMT solver (checking whether the
negation of the implication is unsatisfiable).

\subsection{Example}
Recall the program \textsf{cohendiv} in Figure~\ref{fig:motiv}, which takes as input  two integers $x, y$ and returns the integer $q$ as the quotient of $x$ and $y$.
Given this program and locations of interest \textsf{L1} and
\textsf{L2}, {\tool} automatically discovers the following (loop)
invariants at \textsf{L1}:
\[\begin{array}{l@{\qquad}ll}
x = qy+r  & b = ya&\\
 y \le b & b \le r & \\ 
r \le x & a \le b & 2 \le a + y \\
\end{array}
\]
and the following (postcondition) invariants at \textsf{L2}:
\[
\begin{array}{ll}
x = qy+r & 1 \le q + r\\
r \le x  & r \le y - 1 \qquad 0 \le r 
\end{array}
\]

These equality and inequality relations are sufficiently strong to
understand the function's semantics and verify the correctness of
\textsf{cohendiv}.  More specifically, the nonlinear equation $x=qy+r$
describes the precise behavior of integer division: the dividend $x$
equals the divisor $y$ times the quotient $q$ plus the remainder $r$.
The other inequalities also provide useful information for debugging.  For
example, these invariants reveal several required properties of the
remainder $r$ such as $r$ is non-negative ($r \ge 0$), is at most the
dividend ($r \le x$), but is strictly less than the divisor
($r \le y -1$).
In addition, these invariants can help prove assertions if they exist in the program.
For example, if we want to assert and prove the postcondition stating that the returned quotient is non-negative ($q \ge 0$),
then we can easily do so because the discovered invariants at \textsf{L2} imply $q \ge 0$.\footnote{{\tool} also found this assertion
  and other postconditions at \textsf{L2}, but discarded them because
  they are implied by other discovered invariants and are thus
  redundant.}

As mentioned above, ICE and PIE generate invariants to prove specific assertions.
Thus, given a program with no specific assertion, they will not provide anything useful.
Even when asked to verify a specific assertion, e.g., $x=qy+r$ or other simpler invariants above found by {\tool}, these tools fail to prove them (PIE does not converge and ICE fails to generate invariants to prove the given assertions). We do not have the implementation of the {\it Guess-and-Check} algorithm in~\cite{sharma2013data} to run on this example, however this work does not support inequalities and thus would not generate the inequality invariants shown.

\section{Inferring Polynomial Equalities}
\label{sec:nonlinear_eq}

We now discuss \tool's CEGIR algorithm for generating polynomial
equalities among program variables.
This algorithm integrates the equation solving technique in DIG with KLEE to find invariants.

\subsection{Terms, Templates, and Equation Solving}

{\tool} infers polynomial equalities by searching for solutions to
instantiations of a \emph{template} equation having the form
$c_1t_1 + c_2t_2 \dotsb + c_nt_n = 0$, where $c_i$ are real-valued and $t_i$
are  \emph{terms}. Terms consist of monomials over program variables.
More specifically, given a set $V$ of variables and a
degree $d$, {\tool} creates a set of $n$ terms consisting of
monomials up to degree $d$ from $V$.  For instance, the $n=10$ terms
$\{1,r,y,a,ry,ra,ya,r^2,y^2,a^2\}$ consist of all monomials up to degree
$2$ over the variables $\{r,y,a\}$. 

{\tool} seeks to solve the $c_i$ in the template equation by
\emph{instantiating} the $t_i$ with values observed from
traces. For our example, instantiating the template with the trace
$r=3,y=2,a=6$ would yield the equation 
$c_1 + 3c_2 \dotsb + 36c_n = 0$. If there are $n$ terms, we need at
least $n$ distinct valuations of the variables in $V$. Given the (at
least) $n$ equations that result after instantiation, we solve for the
$c_i$, substituting their solutions into the template to produce
equations over the (combinations of) variables in $V$.

\subsection{Algorithm}

\begin{algorithm}[t]
\small
\DontPrintSemicolon
\SetKwInOut{Input}{input}
\SetKwInOut{Output}{output}

\SetKwFunction{verify}{verify}
\SetKwFunction{infer}{infer}
\SetKwFunction{exec}{exec}
\SetKwFunction{break}{break}
\SetKwFunction{extractVars}{extractVars}
\SetKwFunction{extractEqts}{extractEqts}
\SetKwFunction{createTerms}{createTerms}
\SetKwFunction{createTemplate}{createTemplate}
\SetKwFunction{instantiate}{instantiate}
\SetKwFunction{solve}{solve}

\SetKwData{NotEnoughTraces}{NotEnoughTraces}
\SetKwData{eqts}{eqts}
\SetKwData{template}{template}
\SetKwData{vars}{vars}
\SetKwData{terms}{terms}
\SetKwData{invs}{invs}
\SetKwData{cexInps}{cexInps}
\SetKwData{traces}{traces}
\SetKwData{inps}{inps}
\SetKwData{candidates}{candidates}
\SetKwData{newCandidates}{newCandidates}
\SetKwData{candidate}{candidate}
\SetKwData{candidates}{candidates}
\SetKwData{False}{False}
\SetKwData{True}{True}
\SetKwData{eqts}{eqts}
\SetKwData{sols}{sols}
\SetKwData{eqs}{eqs}

\Input{a program $P$, a location \textsf{L}, a degree $d$}
\Output{polynomial equalities over the variables at \textsf{L} up to degree $d$}
\BlankLine

\vars $\leftarrow \extractVars(P, \textsf{L})$\;
\terms $\leftarrow \createTerms(\vars, d)$\;
\template $\leftarrow \createTemplate(\terms)$\;
\inps, \traces, \eqts, invs $\leftarrow \emptyset, \emptyset, \emptyset, \emptyset$\;

\BlankLine
\While{$|\eqts| < |\terms|$}{
  \cexInps $\leftarrow$ \verify($P, \textsf{L},\False, \inps$)\;
  \If{\cexInps $\equiv \emptyset$}{
    
    \lIf{$\inps \equiv \emptyset$}{ 
      \KwRet $\{\False\}$ //unreachable
    }\lElse{\KwRet \NotEnoughTraces}
  }
  \inps $\leftarrow \inps \cup \cexInps$\;
  \traces $\leftarrow \exec(P, \textsf{L},\cexInps)$\;
  \eqts $\leftarrow \eqts \cup \instantiate(\template, \traces)$
}

\BlankLine
$\sols \leftarrow \solve(\eqts)$\;
$\candidates \leftarrow \extractEqts(\sols, \terms)$

\While{$\candidates \neq \emptyset$}{
  \cexInps $\leftarrow$ \verify($P, \textsf{L},\candidates, \inps$)\;
  \ForEach {\candidate $\in$ \candidates}{
    \lIf{\candidate.stat $\neq$ \False}{\invs.add(\candidate)}
  }

  \lIf{\cexInps $\equiv \emptyset$}{\break}
  \inps $\leftarrow \inps \cup \cexInps$\;
  \traces $\leftarrow$ \exec($P, \textsf{L},\cexInps$)\;
  \eqts $\leftarrow \eqts \cup \instantiate(\template, \traces)$

  $\sols \leftarrow \solve(\eqts)$\;
  $\candidates \leftarrow \extractEqts(\sols, \terms)$

  \candidates $\leftarrow \candidates - \invs$\;
}

\BlankLine
\KwRet \invs \;
\caption{CEGIR algorithm for finding equalities.}
\label{fig:findEqts}
\end{algorithm}

Figure~\ref{fig:findEqts} shows the CEGIR algorithm for finding
polynomial equalities.  Given a program $P$, location \textsf{L}, and a
degree $d$, {\tool} automatically computes all equalities
with degree up to $d$ over the numerical variables at \textsf{L}. (In Section~\ref{sec:results}, we discuss our use of a single parameter that automatically adjusts the degree $d$ depending on the program).

The first steps are to identify the variables at the program location of
interest, and then to construct the terms and template as described
above. Then, in the first loop, we use KLEE to obtain traces to
instantiate the template and thereby produce equations over the
coefficients associated with the generated terms. To obtain traces, we
simply ask KLEE to find cexs producing traces reaching \textsf{L} (more
specifically, the location guarded by $\neg \textsf{False}$ at \textsf{L}).
To avoid getting old inputs, we explicitly ask KLEE to return only new
inputs (by adding assertions that the input variables are not any of
the observed ones). After producing enough equations, we solve them using an off-the-shelf linear equation solver
and extract results representing candidate equality relations among
terms.

Next, the algorithm enters a second loop that iteratively verifies candidate
invariants and obtains cex traces, allowing the inference algorithm to
discard spurious results and generate new invariants. {\tool} accepts
a candidate invariant as long as KLEE cannot find a cex for it within
the timeout period. We repeat the steps of verifying candidate
invariants, obtaining cexs, and inferring new results until we can no
longer find cexs or new results.

Note that unlike the popular CEGAR (counterexample-guided abstract refinement) technique in static
analysis~\cite{clarke2000} that
usually starts with a weak invariant and gradually strengthens it,
\tool's CEGIR algorithm starts with a strong invariant (i.e.,
\textsf{False}) and iteratively weakens it. This is because the
algorithm dynamically infers invariants using observed traces. We
start with few traces and thus likely generate too strong or spurious
invariants. We then accumulate more traces to refute
spurious results and create more general invariants that satisfy all
obtained traces.

We also note that an interesting property of nonlinear polynomial equalities is that they can represent a form of \emph{disjunctive} invariants. For example, $x^2=4$ indicates that $x=2 \vee x=-2$.
In Section~\ref{sec:complexity} we exploit this useful property to find multiple complexity bounds of a program.

\subsection{Example}

We demonstrate this technique by finding the equalities
$b=ya$ and $x=qy+r$ at location \textsf{L1} in the \textsf{cohendiv}
program in Figure~\ref{fig:motiv}, when using degree $d=2$.

For the six variables $\{a,b,q,r,x,y\}$ at \textsf{L1}, together with
$d=2$, we create 28 terms $\{1,a,\dots,y^2\}$. {\tool} uses these
terms to form the template $c_1 + c_2a + \dots c_{28}y^2=0$ with 28
unknown coefficients $c_i$. 
Next, in the first loop, {\tool} uses KLEE to obtain traces such as those given in Figure~\ref{fig:motiv} to form (at least) $28$ equations.
From this set of initial equations, {\tool} solves and extracts seven equalities.

Now {\tool} enters the second loop.
In iteration \#1, KLEE cannot find cexs for two of these candidates $x = qy+r, b = ya$ (which are actually true invariants) and save these as invariants.
KLEE finds cexs for the other five,\footnote{These spurious results often have many terms and large coefficients, e.g., the simplest of these seven is $ry^2 - xy^2 - 72ry + 72xy + 8190q + 1397r - 1397x = 0.$} and {\tool} forms new equations from the cexs.
Next, {\tool} combines the old and new equations to obtain another seven candidates, two of which are the already saved ones (because we also use the old equations).
In iteration \#2, KLEE obtains cexs for the other five candidates.
With the help of the new cex equations, {\tool} now infers three candidates, two of which are the saved ones.
In iteration \#3, {\tool} uses KLEE to find cexs disproving the remaining candidate and again uses the new cexs to infer new candidates.
This time {\tool} only finds the two saved invariants $x = qy+r, b = ya$ and thus stops.

\section{Inferring Octagonal Inequalities}\label{sec:ieqs}

\newcommand{\mytikzscale}[2]{\begin{tikzpicture}[scale={#1}]{#2}\end{tikzpicture}}
\newcommand{\geoshapesBASE}[2]{
  \mytikzscale{#2}{
    \draw[gray!50, step=1cm] (-2.5,-3.5) grid(5.5,2.5);
    {#1}
    \foreach \c in {
      (-2,1),
      (-1,-1),
      (1,-3),
      (3,-2),
      (5,2),
      (2,0)
    }\draw[fill=red,draw=red] \c circle (6pt);
  }
}

\newcommand{\geoshapes}[3]{
  \begin{minipage}{0.17\linewidth}
    \centering
    \geoshapesBASE{#1}{#3}
    \vspace{-0.7cm}
    \caption*{#2}
  \end{minipage}
}

\begin{figure}[h]
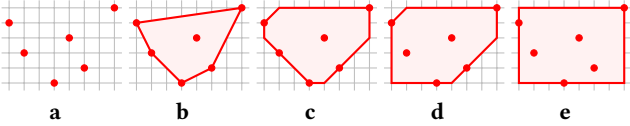

  \hfill
  \geoshapes{}{a}{0.20}
  \hfill
  \geoshapes{
    \draw[mygeoshape] 
    (-2,1) -- (5,2) -- (3,-2) -- (1,-3) -- (-1,-1) -- cycle;
  }{b}{0.20}
  \hfill
  \geoshapes{
    \draw[mygeoshape] 
    (-2,1) -- (-1,2) -- (5,2) -- (5,0) -- (2,-3) -- (1,-3) -- (-2,0) -- cycle;
  }{c}{0.20}
  \hfill
  \geoshapes{
    \draw[mygeoshape] 
    (-2,1) -- (-1,2) -- (5,2) -- (5,0) -- (2,-3) -- (-2,-3) -- cycle;
  }{d}{0.20}
  \hfill
  \geoshapes{
    \draw[mygeoshape] (-2,2) -- (5,2) -- (5,-3) -- (-2,-3) -- cycle;
  }{e}{0.20}
  \hfill
\caption[Geometric Invariant Inference]
{(a) A set of points in 2D and its approximation using the (b) polyhedral, (c) octagonal, (d) zone, and (e) interval regions. These shapes are represented by the conjunctions of inequalities of the forms $c_1v_1+c_2v_2\ge c$, $\pm v_1 \pm v_2 \ge c$, $v_1 - v_2 \ge c$, and $\pm v \ge c$, respectively.}
\label{fig:geoshapes}
\end{figure}

\tool's second algorithm aims to infer linear inequalities among
program variables, essentially by attempting to find a convex polyhedron
in a multi-dimensional space that contains all observed trace points. Figure~\ref{fig:geoshapes}
illustrates several examples of polygons in two-dimensional space.
Figure~\ref{fig:geoshapes}a shows a set of points created from input
traces.
Figures~\ref{fig:geoshapes}b,~\ref{fig:geoshapes}c,~\ref{fig:geoshapes}d,
and~\ref{fig:geoshapes}e approximate the area enclosing these points
using the polygonal, octagonal, zone, and interval shapes that are
represented by conjunctions of \emph{inequalities} of different forms
as shown in Figure~\ref{fig:geoshapes}. These forms of relations are
sorted in decreasing order of expressive power and computational cost.
For example, interval inequalities are less expressive than zone
inequalities, and computing an interval, i.e., the upper and lower
bound of a variable, costs much less than computing the convex hull of
a zone.

{\tool} infers octagonal inequalities. These can be computed
efficiently (linear time complexity) and are also relatively
expressive (e.g. represent zone and interval inequalities as
illustrated in Figure~\ref{fig:geoshapes}). 
Thus, the computation of octagonal inequalities also produces zone and interval inequalities for free.
By balancing
computational cost with expressive power, octagonal relations are
especially useful in practice for detecting bugs in flight-control
software, and performing array bound and memory leak
checks~\cite{CCF05,mine2004weakly}.

\subsection{Terms}

The edges of an inferred octagon are represented by a conjunction of
eight inequalities of the form $a_1v_1 + a_2v_2 \ge k$, where
$v_1,v_2$ are variables, $a_1,a_2\in \{-1,0,1\}$ are 
coefficients, and $k$ is a real-valued constant. For example, from the traces in
Figure~\ref{fig:motiv}, we could infer octagonal inequalities such as
$4 \le r \le 15$ and $3 \le r - y \le 13$ at location \textsf{L1}.

{\tool} infers octagonal inequalities by trying to prove invariants
$t \le k$ for some constant $k$. Here, $t$ is a term
involving two variables so that $t \le k$ is an octagonal constraint,
e.g., $t$ could be $x-y$ or $x+y$. More precisely, we consider all possible terms for $n$ variables: we create $n^2$ variable
pairs from a set of $n$ variables and obtain $8$ octagonal terms
$\{\pm v_1,\pm v_2, \pm v_1 \pm v_2\}$ for each pair $v_1,v_2$. For
each such term we attempt to prove its upper bound $k_1$ and lower
bound $k_2$, if they exist, using the algorithm described next.

\subsection{Algorithm}\label{sec:oct_cegir}
One idea for inferring inequalities would be to iteratively refine
conjectured bounds using cexs, but this can take a long time. For
example, to find the invariant $x \le 100$, we can first infer
$x \le 1$ from traces such as $x\in \{0,1\}$. We can then disprove this
candidate with cexs such as $x\in \{2,3\}$ and weaken the relation to
$x \le 3$, which can also be disproved and weakened. This keeps going
until we get the cex $x=100$, which would allow us to obtain and prove
$x\le 100$. Even worse than taking a long time to reach the bound $k$,
this brute-force approach does not terminate when $x$ has no constant bound. As
such, we use a divide-and-conquer-style search instead.

\begin{algorithm}[t]
\small
\DontPrintSemicolon
\SetKwInOut{Input}{input}
\SetKwInOut{Output}{output}

\SetKwFunction{verify}{verify}
\SetKwFunction{exec}{exec}
\SetKwFunction{ceil}{ceil}
\SetKwFunction{max}{max}

\SetKwData{inps}{inps}
\SetKwData{term}{term}
\SetKwData{minV}{minV}
\SetKwData{maxV}{maxV}
\SetKwData{midV}{midV}
\SetKwData{traces}{traces}
\SetKwData{cexInps}{cexInps}

\SetKwProg{Fn}{Function}{}{}
\SetKwFunction{cub}{\textsc{findUpperBound}}
\Fn(){\cub{\term, \minV, \maxV, $P$, \textsf{L}}}{

\lIf{\minV $\equiv$ \maxV}{\Return \maxV}
\ElseIf{\maxV $-$ \minV $\equiv 1$}{
  \cexInps $\leftarrow$ \verify($P,\textsf{L}, \{\term \le \minV\}, \{\}$)\;
  \lIf{cexInps $\equiv \emptyset$}{\Return \minV}
  \lElse{\Return \maxV}
}
\Else{
  \midV $\leftarrow \lceil\frac{\maxV + \minV}{2}\rceil$\;
  \cexInps $\leftarrow$ \verify($P,\textsf{L}, \{\term \le \midV\}, \{\}$)\;
  \If{cexInps $\equiv \emptyset$}{
    \maxV = midV
  }
  \Else{ //disproved\;
    \traces $\leftarrow$ \exec($P,\textsf{L},\cexInps$)\;
    \minV = \max(\instantiate(term, \traces))\;
  }
  \Return \cub{\term, \minV, \maxV, $P$, \textsf{L}}
}

}
\caption{CEGIR algorithm for finding inequalities.}
\label{ub}
\end{algorithm}

\paragraph{Finding Upper and Lower Bounds.} 
We use the CEGIR algorithm shown in Figure~\ref{ub} to compute a
precise integral upper bound $k$ of a term $t$. 
Similar to a binary search, this algorithm computes $k$ from a given
interval by repeatedly dividing an interval into halves that could
contain $k$. We start with the interval $[minV, maxV]$ where
$maxV= -minV$; our experience is that inequalities are most
useful with small constants, so by default we set $maxV=10$. Next we
check $t\le midV$ where 
$midV = \lceil \frac{maxV + minV}{2} \rceil$. If this inequality is true,
then $k$ is at most $midV$ and thus we reduce the search to the
interval $[minV, midV]$. Otherwise, we obtain counterexample traces
showing that $t > midV$ and reduce the search to $[minV',maxV]$, where
$minV'$ is the largest trace value observed for $t$. Thus this approach
gradually strengthens the guess of $k$ by repeatedly reducing the interval
containing it.

We also use the same approach to find the lower bound of a term
$t$ by computing the upper bound of $-t$. This is possible because
the semantics and results of all computations are reversed when we
consider $-t$. For example, the max over the traces $t \in \{2,3\}$
with respect to $-t$ is $-2$ and $-t \le midV$ indicates the
lower bound of $t$ is at least $-midV$.

The algorithm terminates and gives a precise upper bound value when
$t$ ranges over the integers.
The algorithm stops when $minV$ and
$maxV$ are the same (because we no longer can reduce the intervals) or
when their difference is one (because we cannot compute the exact
$midV$).
Currently {\tool} does not support real-valued bounds.
However, we believe that this algorithm can be extended to handle the case when $t$ ranges over the reals.
More specifically, we can approximate the results by using only whole numbers or values up to certain decimal
places. This sacrifices precision but preserves soundness and
termination, e.g., the invariant is $x \le 4.123$ but we obtain
$x \le 4.2$, which is also an invariant, but less precise.

\subsection{Example}

Recall the program \textsf{cohendiv} from Figure~\ref{fig:motiv}. Suppose
{\tool} wishes to find inequality invariants at \textsf{L1} (within
$[-10,10]$). It first uses KLEE to check
candidate relations $r \le 10, y \le 10,  r  + y \le 10, r -y \le 10, \dots$
and removes those that KLEE refutes. The remaining relations have upper bounds less than or equal to $10$.

For each remaining inequality candidate, {\tool} iterates to find tighter upper
bounds. For example, suppose we wish to find $k$ such that
$r-y \le k$. Since $r-y \le 10$, the algorithm sets
$midV=(10 + -10) \div 2$ which is $0$ and thus tries to check
$r-y \le 0$. This succeeds. However, this turns out to be weaker than
necessary. 
In the next iteration \#2, {\tool} tightens the bound to
$(0 -10) \div 2 = -5$ and checks $r-y \le -5$.
This time KLEE returns a cex showing that
$r-y = -3$. In iteration \#3, {\tool} relaxes the bound to
$(0-3)\div 2 = -1$ and KLEE cannot refute $r-y \le -1$. In 
iteration \#4, {\tool} guesses and checks $(-1-3)\div 2=-2$, in which
case KLEE can find cexs stating that $r-y=-1$. At this point {\tool}
accepts the tightest bound $r-y\le -1$ found in iteration \#3.
The process for finding the lower bounds is similar as described above.

\section{Experimental Results}
\label{sec:results}
{\tool} is implemented in Python and uses the linear equation solver in the SAGE mathematical environment~\cite{sage}.
{\tool} takes as inputs a C program, a list of locations, and interested numerical variables at these locations, and it returns relations among these variables at the considered locations.
As mentioned, {\tool} uses DIG's algorithms to infer invariants and calls the symbolic execution tool KLEE to check results and obtain counterexamples for refinement.
The final step that removes redundant invariants uses the Z3 solver~\cite{z3ms} to check SMT formulas.

We generate numerical invariants of two forms: nonlinear polynomial equations and octagonal inequalities.
For octagonal invariants, {\tool} by default considers the bounds within the range $[-10, 10]$. 
For equalities, {\tool} by default sets a single parameter $\alpha = 200$ so that it can generate invariants without a priori knowledge of specific degrees.
{\tool} automatically adjusts the maximum degree so that the number of generated terms does not exceed $\alpha$.
For example, {\tool} considers equalities up to degree 5 for a program with four variables and equalities up to degree 2 for a program with twelve variables.
We acknowledge that inferring these parameter constants robustly and automatically is important future work.
These constants can be chosen by the {\tool} user; we chose values based on our experience.
Note that the divide and conquer approach to inferring inequalities in Figure~\ref{ub} is quite useful if the user decides to increase the bounds; for range $[-10,10]$ the number of iterations is $\mathsf{log}(20)\approx 5$ rather than $20$ (if we use a brute force algorithm) but for range $[-100,100]$ it is $\mathsf{log}(200)\approx 8$, not $200$ (using brute force).

\paragraph{Experiments.}
We evaluate and compare {\tool} to other invariant analysis systems by considering three experiments.
The first experiment in Section~\ref{sec:program-correctness} determines if {\tool} can discover invariants representing precise semantics and correctness properties of programs having complex arithmetic.
The second experiment in Section~\ref{sec:complexity} explores the use of {\tool}'s invariants to represent precise program complexity bounds.
The last experiment in Section~\ref{sec:comparePIE} compares {\tool}'s performance with the state of the art CEGIR tool PIE.
The experiments reported below were performed on a Linux system with a 10-core Intel i7 CPU and 32 GB of RAM.

\subsection{Analyzing Program Correctness}\label{sec:program-correctness}

\begin{table}[t]
\caption{Results for 27 NLA programs. \checkmark: {\tool} generates sufficiently strong results to prove known invariants.
}
\small
\centering
\begin{tabular}{llc|crr|c}
\textbf{Prog}  & \textbf{Desc}&\textbf{Locs}  & \textbf{V, T, D}     & \textbf{Invs} &\textbf{Time (s)} &\textbf{Correct}                                                                   \\
  \midrule
  cohendiv    &  div         & 2             & 6,3,2    & 11    &  24.57 &   \checkmark \\ 
  divbin      &  div         & 2             & 5,3,2    & 12    &  116.83&   \checkmark \\ 
  manna       &  int div     & 1             & 5,4,2    & 5     &  30.86 &   \checkmark \\  
  hard        &  int div     & 2             & 6,3,2    & 13    &  71.47 &   \checkmark \\ 
  sqrt1       &  sqr root    & 1             & 4,4,2    & 5     &  19.35 &   \checkmark \\ 
  dijkstra    &  sqr root    & 2             & 5,7,3     & 14    &  89.32 &   \checkmark \\ 
  freire1     &  sqr root    & 1             & -        & -     &  -     &   - \\ 
  freire2     &  cubic root  & 1             & -        & -     &  -     &   - \\ 
  cohencu     &  cubic sum   & 1             & 5,5,3     & 5     &  22.56 &   \checkmark \\ 
  egcd1       &  gcd         & 1             & 8,3,2     & 9    &  284.52&   \checkmark \\ 
  egcd2       &  gcd         & 2             & -        & -     &  -     &   - \\ 
  egcd3       &  gcd         & 3             & -        & -     &   -    &   - \\
  prodbin     &  gcd, lcm    & 1             & 5,3,2      & 7     &  45.13 &   \checkmark \\ 
  prod4br     &  gcd, lcm    & 1             & 6,3,3      & 11    &  87.37 &   \checkmark \\ 
  knuth       &  product     & 1             & 8,6,3      & 9     &  84.69 &   \checkmark \\
  fermat1     &  product     & 3             & 5,6,2      & 26    &  185.36&   \checkmark \\
  fermat2     &  divisor     & 1             & 5,6,2      & 8     &  101.83&   \checkmark \\
  lcm1        &  divisor     & 3             & 6,3,2      & 22    &  175.29&   \checkmark \\ 
  lcm2        &  divisor     & 1             & 6,3,2      & 7     &  163.86&   \checkmark \\ 
  geo1        &  geo series  & 1             & 4,4,2      & 7     &  24.41 &   \checkmark \\ 
  geo2        &  geo series  & 1             & 4,4,2      & 9     &  24.33 &   \checkmark \\ 
  geo3        &  geo series  & 1             & 5,4,3      & 7     &  32.38 &   \checkmark \\ 
  ps2         &  pow sum     & 1             & 3,3,2      & 3     &  17.08 &   \checkmark \\ 
  ps3         &  pow sum     & 1             & 3,4,3      & 4     &  17.86 &   \checkmark \\ 
  ps4         &  pow sum     & 1             & 3,4,4      & 4     &  18.55 &   \checkmark \\
  ps5         &  pow sum     & 1             & 3,5,5      & 4     &  19.36 &   \checkmark \\
  ps6         &  pow sum     & 1             & 3,5,6      & 3     &  21.09 &   \checkmark \\
\bottomrule
\end{tabular}
\label{table:results}
\end{table}

\paragraph{Programs.} In this experiment, we focus on generating invariants that capture semantics and correctness properties of programs with nonlinear polynomial invariants.
For this task, we evaluate {\tool} on the NLA~\cite{vuicse2012} test suite consisting of programs involving complex arithmetic.
The suite, shown in Table~\ref{table:results}, consists of 27 programs from various sources collected previously by Rodr\'iguez-Carbonell and Kapur~\cite{1236086,1222597,RCEthesis}.  
These programs are relatively small, on average two loops of 20 lines of code each.  
However, they implement nontrivial mathematical algorithms involving general polynomial properties and are often used to benchmark numerical invariant analysis methods~\cite{RCEthesis}.
To the best of our knowledge, NLA contains the largest number of numerical algorithms with nonlinear polynomial invariants.

Each program in NLA comes with documented or annotated correctness assertions requiring polynomial invariants, mostly loop invariants having nonlinear polynomial equalities.
For evaluation purposes, we consider invariants at the annotated locations and compare them to the documented invariants.

\paragraph{Results.}
Table~\ref{table:results} summarizes the results and reports the medians across 11 runs.
Column \textbf{Locs} gives the number of locations in the programs where we consider invariants.
Column \textbf{Invs} reports the number of equality and inequality invariants discovered by {\tool}.
Column \textbf{V, T, D} shows the number of distinct variables, terms, and the highest polynomial degree in those invariants.
Column \textbf{Time} reports the time in seconds to generate these results, including the time to remove redundant results.
Column \textbf{Correct} indicates whether these invariants matched or were strong enough to prove (imply) the documented invariants.

{\tool} found invariants that matched or were sufficiently strong to prove the documented invariants of 23/27 programs in NLA.
For these programs, we discovered results matched the documented invariants exactly as written in most cases.
{\tool} also achieved invariants that are logically equivalent to the documented ones. 
For example, \lt{sqrt1} has two documented equalities $2a+1=t, (a+1)^2=s$; our results gave $2a+1=t, t^2 + 2t + 1 = 4s$, which is equivalent to $(a+1)^2=s$ by substituting $t$ with $2a+1$.
In many cases, {\tool} also found undocumented invariants, e.g., most of the discovered octagonal inequalities in the \textsf{cohendiv} program in Figure~\ref{fig:motiv} are undocumented.
For \textsf{dijkstra}, {\tool} found the documented invariant describing the semantics of a loop computation, but also discovered an undocumented loop invariant $h^3 = 12hnq - 16npq + hq^2 + 4pq^2 - 12hqr + 16pqr$.
Manual analysis shows that this strange relation is correct and captures detailed dependencies among variables in the loop.
Thus, {\tool}'s strong invariants can help with understanding both \emph{what} the program does and also \emph{how} the program works.
In Section~\ref{sec:complexity}, we further exploit such complex invariants to analyze program complexity.

For these programs, the run time for finding equality invariants is dominated by solving equations because we are solving hundreds of equations with hundreds of unknowns each time.
The run time significantly improves if we restrict the search to invariants up to a certain given degree. 
For example, {\tool} took 2s to find the invariants in \textsf{sqrt1} using degree 2, but it took 20s to find the same invariants using the parameter $\alpha=200$, which queries {\tool} for all invariants up to degree 5 in this program.
For \textsf{egcd1}, the running time is also cut by more than half if we only focus on quadratic invariants.
For inequality invariants, the running time is dominated by checking because we rapidly guess the bound values and check them with KLEE.
Moreover, {\tool} has to perform this ``guess and check'' computation for octagonal constraints over all possible pairs of variables.

We were not able to find invariants for 4/27 programs.
{\tool} was able to infer results matching the documented invariants for \textsf{freire1} and \textsf{freire2}, but KLEE cannot run on these programs because they contain floating point operations. 
For \textsf{egcd2} and \textsf{egcd3}, the underlying SAGE equation solver stopped responding for more than half of the 11 runs (though we observed all correct results for the runs during which the solver worked).
These problems might occur because the solver has to consider hundreds of equations with very large coefficients for hundreds of unknowns.
We are investigating and reporting these problems to the SAGE developers.

\subsection{Analyzing Computational Complexity}\label{sec:complexity}
\begin{figure}[t]
\begin{lstlisting}[numbers=none,mathescape,xleftmargin=0.2cm,emph={L, t}]
void triple(int n, int m, int N){
  assert (0 <= n && 0 <= m && 0 <= N);
  int i = 0, j = 0, k = 0; int t = 0;
  while(i < n){//loop 1
    j = 0; t++;
    while(j<m){//loop 2	       
      j++; k=i; t++;
      while (k<N){k++; t++;}// loop 3		    
      i=k;
    }
    i++;
  }
  [L]
}
\end{lstlisting}
\caption{An example program that has muliple polynomial complexity bounds.}
\label{fig:triple}
\end{figure}

We use {\tool} to discover invariants capturing a program's computational complexity, e.g., $O(n^3)$ where $n$ is some input.
Figure~\ref{fig:triple} shows the program \textsf{triple} with three nested loops, adapted from the program in Figure 2 of Gulwani \emph{et al.}~\cite{pldi09}.
The complexity of this program, i.e., the total number of iterations of all three loops at location \textsf{L}, appears to be $O(nmN)$ at first glance.
Additional analysis yields a more precise bound of $O(n+mn+N)$ because the number of iterations of the innermost loop is bounded by $N$ instead of $nmN$ and it furthermore directly affects the running time of the outermost loop~\cite{pldi09}.

When given this program, {\tool} discovers an interesting and unexpected postcondition at location \textsf{L} about the counter variable $t$, which is a ghost variable introduced to count loop iterations:
\begin{equation*}
\begin{aligned}
N^2mt + Nm^2t - Nmnt - m^2nt - Nmt^2 + mnt^2 + Nmt \\
- Nnt - 2mnt + Nt^2 + mt^2 + nt^2 - t^3 - nt + t^2 = 0.
\end{aligned}
\end{equation*}

At first glance, this quartic (degree 4) equality with 15 terms looks incomprehensible and quite different than the expected bound $O(n+mn+N)$ or even $O(mnN)$.
However, solving this equation for $t$, i.e., finding the roots, yields three solutions $t=0$, $t = N + m + 1$, and $t = n -m(N - n)$.
Careful analysis reveals that these results actually describe three distinct and exact bounds of this program:
\begin{equation*}
\begin{aligned}
&t = 0               &\text{when}\quad& n =   0,\\
&t = N + m + 1       &\text{when}\quad& n \le N,\\
&t = n-m(N - n)      &\text{when}\quad& n >   N.
\end{aligned}
\end{equation*}

Thus, {\tool} can find numerical invariants that represent precise program complexity.
More importantly, the obtained relations can describe expressive and nontrivial \emph{disjunctive} invariants, which capture different possible complexity bounds of a program.

\paragraph{Programs.} We apply {\tool} to find complexity invariants on programs adapted from~\cite{pldi09,Gulwani:2009:SSC:1575060.1575069,DBLP:conf/popl/GulwaniMC09}.\footnote{We disable nondeterministic functions in these programs because currently {\tool} assumes deterministic programs.}
These programs, shown in Table~\ref{table:complexity_results}, are small, but they have nontrivial structures such as nested loops and represent examples drawn from Microsoft's production code~\cite{pldi09}.
For these programs, we introduce the counter variable $t$ and obtain relations among $t$ and other variables, such as inputs, at the program exit locations.

\begin{table}
  \caption{Results for computing programs' complexities.
    \checkmark: {\tool} generates the expected bounds.
  \checkmark \checkmark: {\tool} obtains more informative bounds than reported results.
  \checkmark$^*$: program was slightly modified to assist the analysis.}
\small
\centering
\begin{tabular}{l|ccr|c}
\textbf{Prog}  & \textbf{V, T, D}     & \textbf{Invs} & \textbf{Time (s)} &\textbf{Bound}                                                                   \\
  \midrule
  cav09\_fig1a	& 2,5,2    &   1   & 14.35       & \checkmark \\
  cav09\_fig1d  & 2,5,2    &   1   & 14.24       & \checkmark        \\
  cav09\_fig2d  & 3,2,2    &   3   & 36.09       & \checkmark        \\
  cav09\_fig3a  & 2,2,2    &   3   & 14.24       & \checkmark        \\
  cav09\_fig5b  & 3,5,2    &   5   & 46.88       & \checkmark$^*$        \\
  pldi09\_ex6   & 3,8,3    &   7   & 54.18      & \checkmark         \\ 
  pldi09\_fig2 (triple)& 3,15,4   &   6   & 93.55       & \checkmark\checkmark        \\
  pldi09\_fig4\_1  & 2,3,1    &   3   & 44.26       & \checkmark        \\
  pldi09\_fig4\_2  & 4,4,2    &   5   & 43.72       & \checkmark        \\
  pldi09\_fig4\_3  & 3,3,2    &   3   & 37.54       & \checkmark        \\ 
  pldi09\_fig4\_4  & 5,4,2    &   4   & 56.60       & -         \\
  pldi09\_fig4\_5  & 3,4,2    &   3   & 31.60       & \checkmark$^*$        \\
  popl09\_fig2\_1  & 5,12,3   &   2   & 211.73      & \checkmark\checkmark  \\
  popl09\_fig2\_2  & 4,9,3    &   2   & 65.17       & \checkmark\checkmark        \\
  popl09\_fig3\_4  & 3,4,3    &   4   & 54.70       & \checkmark        \\
  popl09\_fig4\_1  & 3,3,2    &   2   & 42.76       & \checkmark$^*$       \\
  popl09\_fig4\_2  & 5,12,3   &   2   & 158.3       & \checkmark\checkmark         \\
  popl09\_fig4\_3  & 3,3,2    &   5   & 39.28       & \checkmark        \\
  popl09\_fig4\_4  & 3,3,2    &   3   & 34.28       & \checkmark        \\
\bottomrule
\end{tabular}
\label{table:complexity_results}
\end{table}

\paragraph{Results.} Table~\ref{table:complexity_results} shows the median results across 11 runs and has similar format as that of Table~\ref{table:results}.
For column {\bf Bound}, a checkmark denotes that {\tool} generates invariants representing a similar bound to the one reported in the respective paper.
A double checkmark (\checkmark \checkmark) denotes that {\tool} obtains more informative bounds than reported results.
A checkmark with an asterisk (\checkmark$^*$) denotes that the program was modified slightly to assist the analysis. 

As can be seen, {\tool} produced very promising results that capture the precise complexity bounds for these programs.
For 18/19 programs, {\tool} discovered expected or even more informative bounds than reported results in the respective papers. 
For many programs, {\tool} generated equality invariants representing tight bounds, which can be combined with the discovered octagonal inequalities to produce expected bounds. For example, for \textsf{popl09\_fig3\_4}, {\tool} obtained that the number of iterations $t$ is either $n$ or $m$. 
In addition, {\tool} finds inequalities expressing that $t$ is larger than both $n$ and $m$, suggesting that $t$ is equal to $\textsf{max}(n,m)$, which is the bound also obtained in~\cite{DBLP:conf/popl/GulwaniMC09}.
Thus, inequalities, though appearing much weaker compared to the obtained equalities, play an important role to achieve precise program analysis.

Interestingly, in some cases, {\tool} produced results that are more informative than the ones given in the respective papers. 
This is particularly the case for the program $\textsf{triple}$ analyzed earlier because the three distinct bounds produced by {\tool} are strictly less than the bound $n+mn+N$ given in~\cite{pldi09}. 
We note that in most other cases where {\tool} obtained a better bound, the differences were not as apparent as they were for \textsf{triple}.

We performed some adaptations in certain programs to assist the bound analysis.
For \textsf{cav09\_fig5b}, we considered the invariant obtained as one close to the expected bound.
For \textsf{popl09\_fig4\_1}, we inserted an assert statement that $m\ge 0$ at the beginning of the program. 
Finally, for \textsf{pldi09\_fig4\_5}, for the number of iterations $t$ we obtained the three solutions $t=n-m$, $t=m$, or $t=0$, which imply the correct upper bound $\textsf{max}(0, n-m, m)$.

Finally, {\tool} obtained invariants that are not strong enough to show the expected bound for \textsf{pldi09\_fig4\_4}.
However, we would have obtained this bound if we had introduced a variable (or a term) representing the quotient from the division of two other variables in the program. 
In our experiments, when inserting such a variable, we
obtained bounds that were tighter than the ones presented in~\cite{pldi09}. 
Such cases suggest a possible extension to {\tool} for predicting useful terms.

\subsection{Comparing to PIE}\label{sec:comparePIE}
\begin{table}
\small
\caption{Results for the HOLA benchmarks~\cite{Dillig:2013:IIG:2509136.2509511}. \checkmark: Made invariants from PIE. \checkmark \checkmark: Made stronger invariants than PIE. An asterisk $^*$ indicates that verifying the invariant required additional investigation.
$\circ$: Failed to make any invariants, no running time reported in that case.}
\centering
\begin{tabular}{c | c | c | c}
  \textbf{Benchmark} & \textbf{PIE time (s)} & \textbf{{\tool} time (s)} & \textbf{Correct} \\
  \midrule
   01 & 21.88 & 8.75 & \checkmark \checkmark \\
  02 & 36.12 & 10.35 & \checkmark \\
  03 & 56.28 & 108.20 & \checkmark \checkmark \\
  04 & 19.11 & NA & $\circ$ \\
  05 & 25.19 & 13.20 & \checkmark \checkmark \\
  06 & 61.98 & 14.67 & $\checkmark$* \\
  07 & NA & 16.83 & \checkmark \\
  08 & 19.02 & 31.49 & \checkmark \checkmark* \\
  09 & NA & 30.19 & \checkmark \\
  10 & 24.6 & NA & $\circ$ \\
  11 & 27.95 & NA & $\circ$ \\
  12 & 44.52 & NA & $\circ$ \\
  13 & NA & 19.33 & \checkmark \\
  14 & 25.98 & 11.65 & \checkmark \checkmark \\
  15 & 48.30 & 7.7 & \checkmark \\
  16 & 33.19 & 29.07 & \checkmark \\
  17 & 53.36 & 10.33 & \checkmark \checkmark \\
  18 & 21.70 & 7.7 & \checkmark \checkmark \\
  19 & NA & 25.79 & \checkmark \\
  20 & 331.93 & 104.40 & \checkmark \checkmark \\
  21 & 25.65 & 11.60 & $\checkmark$* \\
  22 & 25.40 & 10.90 & \checkmark \\
  23 & 23.40 & 9.07 & \checkmark \checkmark \\
  24 & 51.22 & NA & $\circ$ \\
  25 & NA & 16.76 & \checkmark \\
  26 & 87.64 & 13.50 & \checkmark \\
  27 & 55.41 & 376.80 & \checkmark \\
  28 & 22.16 & NA & $\circ$ \\
  29 & 58.82 & NA & $\circ$ \\
  30 & 33.92 & NA & $\circ$ \\
  31 & 88.10 & 20.39 & $\checkmark$ \\
  32 & 226.73 & NA & $\circ$ \\
  33 & NA & 48.04 & $\checkmark$*  \\
  34 & 121.87 & 12.20 & \checkmark \checkmark \\
  35 & 20.07 & 13.23 & \checkmark \\
  36 & NA & 14.98 & \checkmark \\
  37 & NA & 14.23 & \checkmark\checkmark$^*$ \\
  38 & 37.37 & 10.83 & \checkmark \\
  39 & 24.68 & 2.39 & \checkmark \\
  40 & 60.71 & 17.07 & $\checkmark$* \\
  41 & 34.10  & 15.47 & $\checkmark$ $\checkmark$* \\
  42 & 54.93 & 13.13 & $\checkmark$ $\checkmark$* \\
  43 & 21.16 & 11.3 & \checkmark \\
  44 & 31.92 & 12.3 & \checkmark \\
  45 & 84.00 & 15.3 & \checkmark \\
  46 & 27.56 & NA & $\circ$ \\ 
  \bottomrule
\end{tabular}
\label{table:hola_results}
\end{table}

{\tool} automatically generates invariants for a program location without any
given assertions or postconditions. Other state-of-the-art CEGIR tools such as PIE generate invariants in a goal-directed manner, driven by supplied postconditions. 
In this experiment, we compare {\tool} with PIE's guided inference with postconditions.
This experiment used the HOLA benchmark programs~\cite{Dillig:2013:IIG:2509136.2509511} (adapted by the developers of PIE).
These programs, shown in Table~\ref{table:hola_results}, are short (10-40 LoC each) C programs already annotated with postconditions. 

We first ran PIE on each program and recorded PIE's running time in seconds.
Then, we removed the postcondition and ran {\tool}, asking it to generate invariants at the location in the program where the postcondition was. 
If {\tool} was able to generate invariants, we compared those invariants to the postcondition. 
If the invariants that {\tool} generated were at least precise enough to establish the given postcondition, then 
{\tool} earned a checkmark (\checkmark). If the invariants were more precise,
then {\tool} earned a double checkmark (\checkmark \checkmark). 
For the programs that {\tool} could not generate invariants, then the analysis is assigned the symbol $\circ$. 
The results are in Table~\ref{table:hola_results}.

For 36/46 programs, {\tool} found invariants that were at least as strong as the postconditions in the PIE programs. 
For the remaining 10/46 programs, {\tool} failed to produce the necessary invariants.
For 13 of the 36 programs where {\tool} produced invariants, {\tool} was able to generate stronger invariants. 
For example, for program 17, the target postcondition was $k\ge n$ given a precondition $n\ge 0$, and {\tool} produced, among other invariants, that $k=(n^3 - n + 6)/6$, which implies that for all $n\ge 0$, $k\ge n$.

For programs having the \checkmark$^*$ or \checkmark \checkmark$^*$, {\tool} found stronger invariants that imply the given postcondition, but require additional human effort to reason about. 
For program 42, the given postcondition is $a\ \%\ 2=1$, i.e., $a$ is odd.
{\tool} found the invariants $xy = x + y - 1,  u_1 -a \le -2, a = x + y - 1$, and $2u_1 = x +y - 2$. This set of constraints implies that $x+y=2(u_1+1)$ and $a=x+y-1$, which indicates that $a$ is indeed odd.
But the first invariant in this set produced by {\tool}, also points to another relation among those variables, namely that at least one of $x$ and $y$ is equal to 1, and thus we marked this example with a double checkmark and additionally annotated it with an asterisk.

Another interesting case is with program 8 that contains a postcondition $x<4\ \vee\ y>2$, which has a disjunctive form of strict inequalities that {\tool} does not support.
Instead of generating this, {\tool} returns a stronger relation $x \le y$, which implies this postcondition and therefore proves it.

\paragraph{Summary}
These experiments show that {\tool} is effective in producing expressive and useful invariants.
The NLA experiment in Section~\ref{sec:program-correctness} shows that {\tool} discovers necessary invariants to understand the semantics and check correctness properties of 23/27 NLA programs containing nontrivial arithmetic.
The Complexity experiment in Section~\ref{sec:complexity} indicates that {\tool} discovers useful invariants that capture challenging complexity bounds for 18/19 programs used to
benchmark static complexity analyses. 
We also note that the recent CEGIR tools ICE and PIE cannot find any of
these nonlinear polynomial invariants produced by {\tool} in these experiments, even when we
explicitly tell these tools that they should attempt to verify these
invariants. 
Finally, the HOLA experiment in Section~\ref{sec:comparePIE} shows that {\tool} competes well with PIE and in 36/46 programs discovers invariants that match or are more informative than PIE's.

\subsection{Threats to Validity}\label{sec:threats_validity}
As mentioned earlier, {\tool} can return \emph{unsound} results because KLEE cannot fully verify programs with complex polynomial properties.
We can recover soundness by using a true verifier instead, e.g., 
we are considering the verification tools CPAChecker~\cite{beyer2011cpachecker} and Ultimate Automizer~\cite{heizmann2010nested}, which performed well in the recent SV-COMP 2017~\cite{Beyer2017}.
However, our experience shows that KLEE is effective in finding counterexamples disproving invalid results and thus results that KLEE cannot disprove have high likelihood of being correct.
KLEE is also practical because it can consider challenging invariants that are not understandable to many sound verifiers.

KLEE does not fully support floating point arithmetic and thus {\tool} is limited to finding invariants over integral variables.
KLEE is also language dependent, thus {\tool} considers only C programs.
We are extending {\tool} with additional verification backends that support richer semantics (e.g., arithmetic over the reals) and other languages (e.g., JPF~\cite{jpf} for Java programs).

DIG's algorithms focus on specialized classes of numerical invariants, thus {\tool} is unlikely to find invariants of other, unrelated forms.
However, our results show that {\tool} can often generate invariants that are logically equivalent or sufficiently strong to prove other forms of complex invariants, e.g., disjunctive ones.

Although our benchmark programs have nontrivial structures (e.g., nested loops) with complex arithmetic and have been used to evaluate modern invariant generation systems, these programs are small and do not represent real-world applications containing hundreds of thousands of lines of code.
Nonetheless, we believe that CEGIR is a promising approach to build invariant analysis tools that can scale and handle larger and more complex codebases.
This is because dynamic analysis allows for inferring expressive invariants efficiently from traces and static checkers such as KLEE have become more powerful and practical in recent years.

\section{Related Work}\label{sec:related}
We review related invariant generation techniques using pure static analysis, dynamic analysis, and CEGIR approaches.

\paragraph{Static invariant generation}
Abstract interpretation~\cite{CC76,CC77a,CH78} computes an invariant that over-approximates reachable program states. 
This method starts from a weak invariant representing an initial approximation and iteratively strengthens the invariant by analyzing the structure of the program until reaching a fixed point.
Over-approximation can lead to imprecise information and produce false positive errors.
Thus, major research directions in this area focus on finding \emph{abstract domains} that are sufficiently expressive to retain important information from the programs.
For example, the work in~\cite{CCF05,mine2004weakly} focus on the six-edged zone relations and the eight-edge octagon relations shown in Figure~\ref{fig:geoshapes}.

Rodr\'iguez-Carbonell \emph{et al.}~\cite{1222597,1236086,RCEthesis} use abstract interpretation to generate nonlinear polynomial equalities. 
They first observe that a set of polynomial invariants forms the algebraic structure of an ideal, then compute the invariants using Gr\"obner basis and operations over the ideals, based on the structure of the program until reaching a fixed point. 
The work only analyzes programs with assignments and loop guards that are expressible as polynomial equalities. 
In addition, this technique does not find inequalities and does not support programs with nested loops.

\paragraph{Dynamic invariant generation}
The popular tool Daikon~\cite{ernst2000dynamically,ernst2001dynamically,ernst00quickly,daikon, perkins2004efficient} infers candidate invariants from traces and templates. 
Daikon comes with a large list of invariant templates and tests them against program traces. 
Templates that are violated in any of the test runs are removed and the remainders are presented as the possible invariants. 
For numerical relations, Daikon can find linear relations over at most three variables and has a small number of fixed nonlinear polynomial templates such as $x=y^2$.
In general, the tool has limited support for inequalities and disjunctive invariants.

\paragraph{CEGIR Approaches}
Sharma \emph{et al.}~\cite{sharma2013data} present a {\it Guess-and-Check} technique for inferring equality invariants.
This technique is the standard CEGIR approach, and the ``guess'' component infers equalities using the similar equation solving technique in DIG.
Thus for equality, this technique has the same theoretical power as {\tool}.
The ``check'' component uses the Z3 SMT solver, and in this context, it is interesting to note the various differences in running time caused by the different choices made in the latter and our implementation, and specifically the use of KLEE instead of Z3.
This {\it Guess-and-Check} approach is limited to equality relations and, as mentioned in Section~\ref{sec:oct_cegir}, it is not trivial to extend to finding inequality invariants.

The PIE (Precondition Inference Engine) tool~\cite{Padhi:2016:DPI:2908080.2908099} can generate both preconditions and loop invariants to automatically verify given assertions.
Given an assertion $Q$, the goal is to produce a predicate formula sufficiently strong to ensure the assertion.
To do this, PIE iteratively learns and refines a set of features (predicates over inputs such as $x > 0$) that are sufficiently strong to separate``good'' traces satisfying $Q$ and ``bad'' traces violating $Q$.
These predicates form the required precondition that proves the assertion.
The novelty of PIE is that it does not rely on a fixed class of predicates and can construct necessary predicates during the inference process.
Nonetheless, the tool cannot provide invariants for arbitrary locations in the program, especially if no additional assertions are given. More specifically, on the \textsf{cohendiv} example in Figure~\ref{fig:motiv}, PIE did not converge to an invariant.

The ICE (implication counter-example) learning model~\cite{Garg:2016:LIU:2837614.2837664} is also a CEGIR approach that generates inductive invariants to prove given assertions.
The ``student'' uses a decision learning algorithm to guess candidate invariants expressed over predicates, which separate the good and bad traces. 
The ``teacher'' uses the Boogie verifier to check and provide good, bad, and novel implication counterexamples to help the student infer more precise inductive invariants.
For efficiency, they restrict attention to the octagon domain and search only for predicates that are arbitrary boolean combinations of octagonal inequalities.
Similar to PIE, ICE infers only necessary invariants to prove assertions.
Even when provided with assertions such as the postconditions of the program \textsf{cohendiv} in Figure~\ref{fig:motiv}, ICE fails to prove them.
We note that part of the reason might be because ICE does not support arithmetic operations such as division and modulo.

\section{Conclusion}
We present {\tool}, a CEGIR-based tool that discovers numerical invariants at arbitrary program locations.
{\tool} uses a dynamic analysis to infer invariants and the test-input generation tool KLEE to verify them.
For invalid invariants, KLEE returns counterexample traces that are then used to help the inference algorithm discard invalid results and to find new invariants.
The use of KLEE allows {\tool} to work on programs with nontrivial arithmetic and discover useful and complex invariants.
Preliminary experiments show that {\tool} often outperforms state-of-the-art CEGIR systems in discovering invariants required to understand and analyze semantics, correctness, and complexity properties of programs.

\begin{acks}
We thank the anonymous reviewers for their detailed feedback and
helpful comments.
This research was supported by DARPA under contracts
FA8750-15-2-0104 and FA8750-16-C-0022.
\end{acks}

\newpage
\bibliographystyle{ACM-Reference-Format}
\balance
\bibliography{paper} 
\end{document}